# A simple technique for evaluating dipole moments of Bloch states in tetrahedral semiconductors.

Jacob B Khurgin

*Dept. of ECE, Johns Hopkins University Baltimore MD 21218*

*jakek@jhu.edu*

**Abstract:** Permanent dipole moments of electronics states in non-centro-symmetric materials play pivotal role in many phenomena. Correctly evaluating them presents an arduous task and usually requires full knowledge of the band structure as well as understanding the intricate concepts of Berry curvature. Here we show that in a few cases (e.g. zinc blende and wurtzite) a rather facile first-principle analytical derivation of the permanent dipole moments using L'Hôpital's rule can be performed, and the values and dispersion of these dipoles near high symmetry points can be found using just a couple of widely available material parameters. The results will hopefully contribute to better understanding of shift currents, optical rectification and other electro-optical phenomena.

Permanent (diagonal) electric dipole moments $\mathbf{d}_i = -e\langle\psi_i|\mathbf{r}|\psi_i\rangle$ of states in condensed matter play defining role in determining many important properties of materials lacking inversion symmetry, such as ferro-, pyro-, and piezo-electricity, linear electro-optic (Pockels) effect and second nonlinear optical properties[1-3]. While the concept of a permanent dipole associated with the asymmetric distribution of charge is easy to visualize, actually evaluating it rigorously is problematic due to translational symmetry, as one is faced with ambiguity of choosing the unit cell where the dipole moment is supposed to be evaluated, as well as with the choice of the boundary conditions.

Permanent dipole of polar bond states can be introduced phenomenologically, by assigning dipoles proportional to polarity to each of the tetrahedral bonds. Based on Phillips-Van Vechten model[4, 5], this transparent "bond charge" approach[6, 7], while somewhat lacking in rigor, has been very successfully used to predict the strength of second order susceptibility[8], Pockels effect[9], optical rectification (OR)[10] as well as linear electro absorption[11] without reverting to full band structure calculations, as in, for instance [12, 13]

The permanent dipole conundrum is rigorously resolved by finding a "relative" dipole moment[14], i.e. comparing the dipoles of similar crystalline structures with and without center of inversion, for instance GaAs (zinc blende) and Ge (diamond). In their pioneering work King-Smith and Vanderbilt [15] identified in permanent polarization a geometric Berry[16]–Pancharatnam [17] quantum phase and succeeded in evaluating piezoelectric coefficient of GaAs. The theory was later successfully applied to evaluation of piezoelectricity [18, 19] and ferroelectricity [20, 21] II-VI The method has been conveniently summarized in works by Resta [22, 23], where connection to the Zak's phase [24] had also been made.

A key feature of the aforementioned graceful method is evaluation of the expected value of coordinate $\langle \mathbf{r_k} \rangle = \langle \psi_\mathbf{k} | \partial / \partial \mathbf{k} | \psi_\mathbf{k} \rangle$ [25] , related to Berry connection. It requires full knowledge of the dispersion over the entire Brillouin zone (BZ) – that dispersion can only be obtained from diagonalizing full Hamiltonian with many matrix elements[26, 27]. Some of these matrix elements arise from the lack of inversion symmetry, and it is due to these matrix elements that non-zero $\langle \mathbf{r_k} \rangle$ arises. This appears to be a rather longwinded way to arrive at the end result – first matrix elements (often experimentally determined rather than derived) are used to obtain the high order in **k** dispersion of the composition of Bloch states, and only then one can evaluate the expected value of the coordinates and dipole moment. Furthermore, in order to analyze the non-resonant properties of non-centrosymmetric materials it should not be necessary to determine contributions of each and every state in BZ with subsequent integration, as one can use a simple bond orbital model[5] to gain a good grasp of the strength of a given phenomenon, be it a piezoelectric effect, or Pockels effect and Second Harmonic Generation in the transparency region well below the bandgap. This proven approach is justified by the fact that the highest density of states is located near the edges of BZ zone, so one can first perform integration of charge density of all Bloch states and then use the properties of the states near the special points (e.g. *X* points for zinc blende lattice) as "average" properties. Thus, for instance, linear optical properties are well described using a simple two-level Lorentz model with resonant energy taken between conduction (CB) and valence bands (VB) near $X$ point – the so called Penn gap[28]. On the other hand, resonant phenomena associated with non-centrosymmetric media depend only on states near the fundamental bandgap ($\Gamma$ point) that are simple combinations of a very few basis states. If one could estimate permanent dipoles of those simple combinations, one would avoid the intricate calculations.

The aforementioned resonant phenomena include electro-absorption and resonant second order nonlinearities, in particular optical rectification (OR) [29] and closely related shift current (SC)[30]. In the latter phenomena , important for THz generation [10], and possibly for photovoltaics[31], the strength of photogenerated polarization (OR) or its derivative (SC) is proportional to the difference between permanent dipoles of Bloch states in VB and Conduction band (CB) $\Delta \mathbf{d}_{cv}(\mathbf{k}) = \mathbf{d}_{cc}(\mathbf{k}) - \mathbf{d}_{vv}(\mathbf{k})$. Since absorption strength depends on the interband dipole $\mathbf{d}_{cv}(\mathbf{k}) = -e\langle \psi_{c\mathbf{k}}|\mathbf{r}|\psi_{v\mathbf{k}}\rangle$, each pair of CB and VB states with momentum $\mathbf{k}$ (neglecting photon momentum) generates polarization (OR) and SC proportional to the tensor $\mathbf{R}(\mathbf{k}) = \Delta \mathbf{d}_{cv}(\mathbf{k}) \otimes \mathbf{d}_{cv}(\mathbf{k}) \otimes \mathbf{d}_{vc}(\mathbf{k})$. The interband dipole can be evaluated via the usual relation $\mathbf{d}_{cv} = -e\hbar \mathbf{P}_{cv} / m_0 (\mathrm{E}_c - \mathrm{E}_v)$, where the momentum matrix element [26] is roughly within the same order of magnitude $P_{cv} \sim \hbar / a_b$, where $a_b$ is the bond length, for all tetrahedral semiconductors[32, 33] and many other materials. No such simple relation between dipole and momentum appears to exist for the "diagonal" matrix elements of dipole and momentum because $P_{cc}(P_{vv}) = 0$, energy difference also obviously being equal to zero.

Nonetheless, as we show below, in certain cases (e.g. tetrahedral covalent lattices, like zinc blende or wurtzite) one can still benefit from the momentum-dipole relation and resolve the "0/0" uncertainty by invoking the well-aged L'Hôpital's rule [34]. Using this technique allows rigorous, straightforward and unambiguous evaluation of differences between the dipoles of Bloch states, their values simply related to the effective masses and bond polarities. It then permits one to estimate OR and SC as a function of excitation energy as well as electronic contributions to Pockels effect and (linear) electro absorption, which, hopefully, indicates that this work has some practical value beyond being a scientific curiosity.

We start with the standard picture[35] of a zinc blende semiconductor (wurtzite can be treated similarly) with 8 basis $s$ and $p$ states of cation and anion atoms (Fig.1): $S_{A(C)}, X_{A(C)}, Y_{A(C)}, Z_{A(C)}$. Near BZ center $\Gamma$ the states get combined into 8 new basis states, 4 bonding (VB) and 4 anti-bonding (CB) ones.

$$\begin{aligned} S_{v(c)} &= 2^{-1/2} N^{-1/2} \sum_{i=1}^{N} \left( \sqrt{1 \pm \alpha_s} S_{A,i} \pm \sqrt{1 \mp \alpha_s} S_{A,i} \right) \\ X_{v(c)} &= 2^{-1/2} N^{-1/2} \sum_{i=1}^{N} \left( \sqrt{1 \pm \alpha_p} X_{A,i} \mp \sqrt{1 \mp \alpha_p} X_{C,i} \right) \\ Y_{v(c)} &= 2^{-1/2} N^{-1/2} \sum_{i=1}^{N} \left( \sqrt{1 \pm \alpha_p} Y_{A,i} \mp \sqrt{1 \mp \alpha_p} Y_{C,i} \right) \\ Z_{v(c)} &= 2^{-1/2} N^{-1/2} \sum_{i=1}^{N} \left( \sqrt{1 \pm \alpha_p} Z_{A,i} \mp \sqrt{1 \mp \alpha_p} Z_{C,i} \right), \end{aligned} \quad (1)$$

summation is taken over all unit cells, $\alpha_{s(p)}$ is the polarity that is equal to

$$\alpha_{s(p)} = \Delta E_{s(p)} / \sqrt{\Delta E_{s(p)}^2 + V_{ss(pp)}^2} = \Delta E_{s(p)} / E_{b,s(p)}, \quad (2)$$

where $V_{ss(pp)} = \langle S(X)_A | H | S(X)_C \rangle$ and $\Delta E_{s(p)} = (E_{s(p)C} - E_{s(p)A})/2$ are covalent and ionic energies of the polar bonds, and $E_{b,s(p)} = \sqrt{\Delta E_{s(p)}^2 + V_{ss(pp)}^2}$ is the total bonding energy. Note that at this point we are not yet considering effect of spin-orbit (SO) interaction and focus on the triple degenerate bonding states in the VB $X(Y,Z)_v$ shown in Fig.1a, with energy

$$E_{vp} = (E_{pC} + E_{pA})/2 - E_{b,p} \quad (3)$$

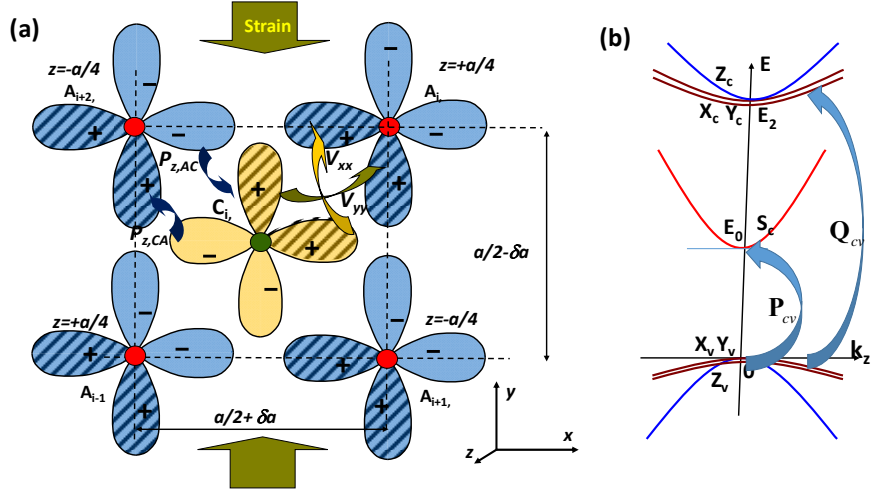

**Figure 1.** (a) Bonding p-orbitals in zinc blende lattice with relevant covalent energies $V_{xx(yy)}$ and inter-site momentum matrix elements $P_z$. When uniaxial strain is applied the degeneracy of $X$ and $Y$ states is lifted. (b) Sketch of the band structure near BZ center in the absence of SO interactions.

As a first step, we introduce the matrix element of momentum between two states on neighboring sites,

$$P_{z,CA} = N^{-1} \sum_{i=1}^{N} \sum_{j=1}^{4} \langle X_{C,i} | p_z | Y_{A,ij} \rangle = 4N^{-1} \sum_{i=1}^{N} \langle X_{C,i} | p_z | Y_{A,i1} \rangle, \quad (4)$$

where simple analysis of Fig.1 reveals that all four terms corresponding to the nearest neighbors have the same sign. At the same time, the matrix element

$$P_{z,AC} = N^{-1} \sum_{i=1}^{N} \sum_{j=1}^{4} \langle X_{A,i} | p_z | Y_{C,ij} \rangle = -P_{z,CA}, \quad (5)$$

since the sign of derivative $dY/dz$ changes between cation and anion. Subsequently, when one evaluates the momentum matrix element between two degenerate VB states (1) cancellation ensues.

$$\langle X_v | p_z | Y_v \rangle = \frac{1}{2N} \sum_{i,j=1}^{N} \left( \sqrt{1+\alpha_p} X_{A,i} - \sqrt{1-\alpha_p} X_{C,i} \right) p_z \left( \sqrt{1+\alpha_p} Y_{A,j} - \sqrt{1-\alpha_p} Y_{C,j} \right) = \\ \frac{1}{2N} \sqrt{1-\alpha_p^2} \sum_{i=1}^{N} \sum_{j=1}^{4} \langle X_{A,i} | p_z | Y_{C,ij} \rangle + \frac{1}{2N} \sqrt{1-\alpha_p^2} \sum_{i=1}^{N} \sum_{j=1}^{4} \langle X_{C,i} | p_z | Y_{A,ij} \rangle = 0 \quad (6)$$

As expected, momentum matrix element between two degenerate states is equal to zero, and if we are to use relation between the momentum and dipole matrix elements we will end up with the aforementioned "0/0" uncertainty. To resolve this uncertainty we consider a small perturbation that can lift the degeneracy, for instance a uniaxial strain that changes the lattice constant so that $a_{x(y)} = a \pm \delta a$ as shown in Fig.1a. Accordingly, as the interatomic distances change, covalent bonding energies also change as $V_{xx(yy)} = V_{pp} \mp \delta V$ and the degeneracy gets lifted (as long as $V_{pp} \neq 0$)

$$E_{X(Y)} = E_0 \mp V_{pp} / \sqrt{\delta E_p^2 + V_{pp}^2} \, \delta V = E_0 \mp \sqrt{1-\alpha_p^2} \, \delta V, \quad (7)$$

while the polarity changes as well

$$\alpha_{x(y)} = \alpha_p \mp \frac{\bar{V}_{xx} \Delta E_p}{\left( \Delta E_p^2 + V_{pp}^2 \right)^{3/2}} \delta V = \alpha_p \mp \alpha_p (1-\alpha_p^2) \delta V / V_{pp} \quad (8)$$

Now, re-calculation the matrix element of momentum (6) leads to a non-vanishing result

$$\langle X_v | p_z | Y_v \rangle = \frac{1}{2N} \sum_{i,j=1}^{N} \left( \sqrt{1+\alpha_x} X_{A,i} - \sqrt{1-\alpha_x} X_{C,i} \right) p_z \left( \sqrt{1+\alpha_y} Y_{A,j} - \sqrt{1-\alpha_y} Y_{C,j} \right) =$$
$$\frac{1}{2N}\sqrt{(1+\alpha_x)(1-\alpha_y)} \sum_{i=1}^{N}\sum_{j=1}^{4} \langle X_{A,i} | p_z | Y_{C,ij} \rangle + \frac{1}{2N}\sqrt{(1+\alpha_x)(1-\alpha_y)} \sum_{i=1}^{N}\sum_{j=1}^{4} \langle X_{C,i} | p_z | Y_{A,ij} \rangle = \quad (9)$$
$$\frac{1}{2} P_{z,CA} \left[ \sqrt{(1+\alpha_x)(1-\alpha_y)} - \sqrt{(1-\alpha_x)(1+\alpha_y)} \right] = \frac{1}{2} P_{z,CA} \frac{\alpha_x - \alpha_y}{\sqrt{1-\alpha_p^2}} = -\alpha_p P_{z,CA} \sqrt{1-\alpha_p^2} \delta V / V_{pp}$$

From (7) and (9) we now obtain, cancelling $\delta V$ in numerator and denominator in full accordance with the L'Hopital's rule,

$$z_{xy,v} = \langle X_v | z | Y_v \rangle = \hbar \langle X_v | p_z | Y_v \rangle / m_0 (E_Y - E_X) = \alpha_p \hbar P_{z,CA} / 2 m_0 V_{pp} = \alpha_p \hbar P_{z,CA} / 2\sqrt{1-\alpha_p^2} m_0 E_{b,p} \quad (10)$$

Note that the last expression does not diverge, since if and when the covalent energy becomes small ($\alpha_p \to 1$) the overlap between wavefunctions on neighboring sites also becomes very small and so does $P_{z,CA}$. At any rate, that is a rather hypothetical case because in real zinc blende and wurtzite materials polarity does not exceed 0.8 [36]

To evaluate matrix element (10) consider the momentum matrix element between bonding and antibonding $p$-states, i.e. between the VB and upper CB as shown in Fig.1b

$$\langle X_v | p_z | Y_c \rangle = \frac{1}{2N} \sum_{i,j=1}^{N} \left( \sqrt{1+\alpha_x} X_{A,i} - \sqrt{1-\alpha_x} X_{C,i} \right) p_z \left( \sqrt{1-\alpha_x} Y_{A,j} + \sqrt{1+\alpha_x} Y_{C,j} \right) =$$
$$\frac{1}{2N}(1+\alpha) \sum_{i=1}^{N}\sum_{j=1}^{4} \langle X_{A,i} | p_z | Y_{C,ij} \rangle - \frac{1}{2N}(1-\alpha) \sum_{i=1}^{N}\sum_{j=1}^{4} \langle X_{C,i} | p_z | Y_{A,ij} \rangle = P_{z,CA} = Q_{cv} \quad (11)$$

According to the $\mathbf{k} \cdot \mathbf{p}$ theory [27] one can relate the momentum matrix element $Q_{cv}$ and the heavy hole effective mass $m_0 / m_{hh} \approx 1 - 2Q_{cv}^2 / m_0 E_2$, where $E_2 \approx 2E_{b,p}$ is the splitting between the bonding and antibonding $p$ states at $\Gamma$ point that is not far from the value of bandgap at $X$ point of BZ, which itself is close to the Penn gap value [37]. Then one obtains from (10) and **Error! Reference source not found.**

$$z_{xy,v} \approx \frac{\alpha_p}{\sqrt{1-\alpha_p^2}} \sqrt{\frac{\hbar^2 (1 + m_0 / |m_{hh}|)}{2 m_0 E_2}} \quad (12)$$

For III-V zinc blende semiconductors, such as GaAs, $m_{hh} \sim 0.5 m_0$ and $E_2 \sim 5 eV$, so that $z \sim 1.4 Å$. Note that $1.4 Å$ is not far from the distance between anion and cation, $a/4$, which lends credibility to this result. Also note that in more polar, III-N and II-VI[35] materials the effective mass of hole is larger $m_{hh} \sim 1 - 1.5 m_0$ and the $E_2$ gap is also larger by a few eV, but at the same time covalence $\sqrt{1-\alpha_p^2}$ is smaller so in the end the estimate $z_{xy,v} \sim \alpha_p a_b / 4$ always gives correct magnitude.

We can now apply the result (12) to the study of the phenomena mentioned earlier. First let us consider the resonant phenomena, such as OR and SC in the direct bandgap materials, so only the states near the fundamental bandgap need to be considered For that rotate the basis by 45 degrees around z axis and introduce new state $X'_v(Y'_v) = 2^{-1/2}(X_v \pm Y_v)$ as shown in Fig.2a. While the permanent dipole moments of these new states are ill defined, as mentioned above due to ambiguity in choosing the origin of coordinates, the all-important difference of the dipoles can be found unambiguously as $\Delta \mathbf{d}_{x'y',v} = -e\hat{\mathbf{z}}(\langle X'_v|z|X'_v\rangle - \langle Y'_v|z|Y'_v\rangle) = -2e\hat{\mathbf{z}}\langle X_v|z|Y_v\rangle = -2ez_{xy,v}\hat{\mathbf{z}}$. Similarly, evaluating the difference between dipoles of band edge states in VB and CB yields $\Delta \mathbf{d}_{cx'} = -e\hat{\mathbf{z}}(\langle X'_v|z|X'_v\rangle - \langle S_c|z|S_c\rangle) = -e\hat{\mathbf{z}}z_{xy,v}$ and $\Delta \mathbf{d}_{cy'} = e\hat{\mathbf{z}}z_{xy,v}$. Since, as mentioned above, only the difference between the dipoles matters, it is convenient to choose $\langle S_c|z|S_c\rangle = 0$. Of course, this result can be easily interpreted – the $X'_v(Y'_v)$ state is located in the (110) ($(1\bar{1}0)$) plane, hence, as shown in Fig.2a, strong polar σ bonds are formed only where the anion atom is above (below) cation atom, and therefore the center of charge is shifted "up" ("down") by $\pm z_{xy,v}$.

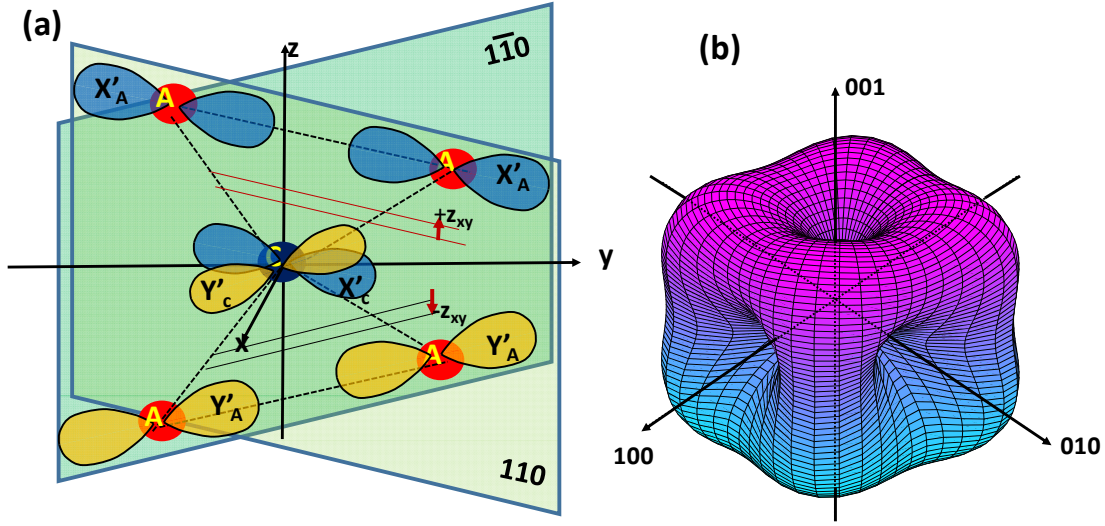

**Figure 2.** (a) VB states $X'_v$ and $Y'_v$ near BZ center using basis rotated by 45 degrees around z axis. $X'_C(Y'_C)$ state makes strong covalent σ bond only with $X'_A(Y'_A)$ state above (below) it. (b) absolute value of permanent dipole of VB HH and LH states versus direction of k-vector.

When the light polarized in the $(110)$ $((1\bar{1}0))$ plane (i.e. having both *x* and *y* components) interacts with the material, only the electrons residing in $X'_v(Y'_v)$ states will make transition into the $S_c$ state leaving behind holes whose charge will have dipole $\mp \Delta \mathbf{d}_{cx'}$ along *z* direction[10]. Therefore, the OR and SC tensors will have non zero $r_{zxy} \equiv r_{36} = r_{14} = r_{25}$ components[10, 29, 38]. Furthermore, applying DC field along *z* direction will shift the energies of $X'_v(Y'_v)$ states by $\Delta E_{x'y'} = \mp e z_{xy,v} E_{DC}$ inducing anisotropy for the light polarized in $(1\bar{1}0)$ and $(110)$ planes. The anisotropy of refractive index is of course manifested by the Pockets coefficient $r_{14}$ [9], but, what is perhaps less widely known, there is also a small change in absorption as band edges for the different orbitals shift in the opposite directions[39]. Obviously applying the strain along, say $(1\bar{1}0)$ axis would shift the charge center of $Y'_v$ states without affecting $X'_v$ states thus generating piezoelectric fields.

Now, when SO interaction is taken into account, the heavy, light, and split-off hole states with wave vector in the [110] and [1$\bar{1}$0] directions become [40]

$$HH_{110(1\bar{1}0)} = (1/2)^{1/2}[Y_v'(X_v') + iZ_v']$$
$$LH_{110(1\bar{1}0)} = (1/6)^{1/2}[Y_v'(X_v') - iZ_v']\uparrow + (2/3)^{1/2} X_v'(Y_v')\downarrow \quad (13)$$
$$SO_{110(1\bar{1}0)} = (1/3)^{1/2}[Y_v'(X_v') - iZ_v']\uparrow + (1/3)^{1/2} X_v'(Y_v')\downarrow$$

Therefore, the dipoles become $z_{[110])([1\bar{1}0]}^{HH} = \mp z_{xy,v}/2$ and $z_{[110])([1\bar{1}0]}^{LH} = \pm z_{xy,v}/2$, and thus heavy hole and light holes states have opposite permanent dipoles, while split-off band has none. Of course, since different states have different energies and densities of states, the dipoles do not cancel each other and such resonant phenomena as OR and SC become possible. For an arbitrary direction of the wave-vector **k** defined by polar and azimuthal angles $(\theta, \varphi)$ one obtains following expression for the dipole moment of the heavy hole state near BZ center,

$$\mathbf{d}_{hh}(\mathbf{k}) = \frac{1}{2}ez_{xy,v}\begin{bmatrix} \sin 2\theta \sin\varphi \\ \sin 2\theta \cos\varphi \\ -\sin^2\theta \sin 2\varphi \end{bmatrix} = -\mathbf{d}_{lh}(\mathbf{k}) \quad (14)$$

The magnitude of dipole moment

$$d_{hh}(\mathbf{k}) = \frac{1}{2}ez_{xy}\sqrt{\sin^2(2\theta) + \sin^2(2\varphi)\sin^4(\theta)} \quad (15)$$

is plotted in Fig.2b that reveals the $\bar{4}3m$ symmetry of zinc blende lattice with maxima pointing in the [111] directions corresponding to tetrahedral bonds. Obviously, integration over the wavevector directions for a given energy $E_{hh}$ ends up with 0. This has a very important consequence – intraband relaxation within BZ occurring on the sup-picosecond scale[41] quickly depletes the photo-generated polarization following the optical pulse, thus OR and SC always occur on ultrafast scale even though real carriers are excited by above the bandgap light. SC density can be evaluated as $\mathbf{J}_{sh} = \eta\mathbf{E}\mathbf{E}^*$, where the injection efficiency tensor has only $\eta_{14} = \eta_{25} = \eta_{36}$ components. For the light polarized in one of the $[1\bar{1}0]$ direction one can obtain after integrating over the directions of wavevector, $J = e^*I/\hbar\omega$, where $I$ is the light power density the "effective shift charge" can be found as $e^* = (2/15)\alpha z_{xy,v}e$, where $\alpha$ is the absorption coefficient. With $\alpha \sim 10^3 - 10^4 cm^{-1}$ near bandgap one obtains $e^*/e \sim 10^{-6} - 10^{-5}$, or shift current sensitivity of a few $\mu A/W$, similar to observed in [30] which is many order of magnitude less than in a photodiode, but, at the same time SC does lead to a sub-picosecond response unattainable in a photodiode. The resonant OR induced field can then be estimated as $E_{OR} = e^* I \tau_{ib} / \varepsilon_0 \varepsilon \hbar \omega$, where $\tau_{ib} \sim 10^{-13} s$ is the aforementioned intraband relaxation time, or $E_{OR}/I \sim 10^{-7} - 10^{-6} V \cdot cm/W$. Therefore, with $I \sim GW/cm^2$ rectification fields of a few $KV/cm$ can be induced with the above the bandgap excitation in line with experiment in [29].

When it comes to a non-resonant response, say SHG or Pockels effect, one can consider hybrid bond model, where one can easily show that each hybrid bonding orbital $h_v = \frac{1}{2}(S_v \pm X_v \pm Y_v \pm Z_v)$ has dipole moment that is directed along the bond (i.e. one of the [111] directions and has magnitude $d_h = (\sqrt{3}/2)ez_{xy,v}$. The bond model implicitly incorporates the states throughout the entire Brillouin zone and is valid as long as one is far from absorption edge. Then one can revert to the well-established bond charge model [6, 8, 42], but this time with the off-center location of the bond charged (i.e. dipole) established rigorously, rather than simply phenomenologically introduced[43]. Obviously, since the bond dipoles values and their dependence on bond polarity estimated here are close to the one introduced in the aforementioned works, all the results obtained in those works remain valid.

In conclusion, a simple analytical path for determining permanent dipoles (or strictly speaking, the differences between them) of Bloch states in tetrahedral lattices has been identified. This accessible and straightforward technique uses L'Hopital's rule and requires knowledge of just a few well known parameters - bond polarities, effective masses, and bandgaps near the high symmetry points. Using this method, a very transparent picture emerges of several important phenomena observed in these lattices, such as SC, OR, second order nonlinear processes, Pockels effect, etc., and the magnitudes of these effects are easily ascertained. The method can in principle be applied to a number of lattices with degenerate valence states, such as, for instance in 2D materials. Being intuitively anticipated, i.e. short of surprises, these results clearly do not break new fundamental grounds, nor do they portend an impending arrival of ultra-efficient and hyper-sensitive nanoscale devices, as one would have wished. Nevertheless, community may find some intrinsic (at least pedagogical) value in this elegant (admittedly in my very subjective view) alternative way to resolve a long standing issue that has baffled the community for quite some time.